\documentclass[aps,prl,twocolumn,groupedaddress,showpacs]{revtex4-1}
\usepackage{graphicx}
\usepackage{epstopdf}
\newcommand{\euc}{EuC$_2$ }

\begin{document}

\title{Measurements of thermodynamic and transport properties of EuC$_2$: \\
a low-temperature analogue of EuO}

\author{O.~Heyer$^1$, P.~Link$^1$, D.~Wandner$^1$, U.~Ruschewitz$^1$, and T.~Lorenz$^1$}
\email[]{tl@ph2.uni-koeln.de}
\affiliation{$^1$II.\ Physikalisches Institut, Universit\"{a}t zu
K\"{o}ln, Z\"{u}lpicher Stra{\ss}e 77, 50937 K\"{o}ln, Germany \\
$^2$Institut f\"{u}r Anorganische Chemie, Universit\"{a}t zu
K\"{o}ln, Greinstra{\ss}e 6, 50939 K\"{o}ln, Germany}

\date{\today}

\begin{abstract}
EuC$_2$ is a ferromagnet with a Curie-temperature of $T_C \simeq 15\,$K. It is semiconducting with the particularity that the resistivity drops by about 5 orders of magnitude on cooling through $T_C$, which is therefore called a metal-insulator transition.   
In this paper we study the magnetization, specific heat, thermal expansion, and the resistivity around this ferromagnetic transition on high-quality
EuC$_2$ samples. 
At $T_C$ we observe well defined anomalies in the specific heat $c_p(T)$ and thermal expansion $\alpha(T)$ data. The magnetic contributions of
$c_p(T)$ and $\alpha(T)$ can satisfactorily be described within a mean-field theory, taking into account the magnetization data.  In zero
magnetic field the magnetic contributions of the specific heat and thermal expansion fulfill a Gr\"uneisen-scaling, which is not preserved in finite
fields. From an estimation of the pressure dependence of $T_C$ via Ehrenfest's relation, we expect a considerable increase of
$T_C$  under applied pressure due to a strong spin-lattice coupling. 
Furthermore the influence of weak off stoichiometries $\delta$ in EuC$_{2 \pm \delta}$ was studied. 
It is found that $\delta$ strongly affects the resistivity, but hardly changes the transition temperature. In all these aspects, the 
behavior of EuC$_2$ strongly resembles that of EuO.
\end{abstract}



\maketitle

\section{Introduction}
Metal dicarbides are composed of a metal ion ($M^{n+}$) and
an acetylide ion (C$_2^{2-}$).
Depending on the metal their physical properties may vary over a wide range 
and thus these systems are investigated since several decades. The electronic and magnetic properties 
depend on the valence of the metal ion and these compounds can be insulators, metals or superconductors  \cite{gulden97}.
Within a simple model which fits for alkaline-earth as well as rare-earth dicarbides \cite{long92a,long92b}, trivalent or
quadrivalent metals form metallic dicarbides, while divalent
metals form insulating compounds. Of particular interest are therefore metals which can realize both, the
divalent as well as the trivalent state as, e.g., Eu and Yb. Experimental studies of these systems are however rare, what is most probably related to
the fact that both EuC$_{2}$ as well as YbC$_{2}$ very rapidly decompose in air. Since 1964 several publications have reported the synthesis of
EuC$_{2}$, but still there is
a disagreement if it crystallizes in a tetragonal \cite{gebelt64,sakai82}
or a monoclinic \cite{faircloth68, wandner10} structure at room
temperature and below. The tetragonal
modification of EuC$_{2}$ and its solid solutions with La and Gd 
have been investigated in Ref.~\onlinecite{sakai82} and it was found that at least some of the
results are not consistent with the simple model mentioned above. For example, the
magnetization data of EuC$_{2}$ reveal a ferromagnetic transition at
$T_{C} \simeq 20$K with a saturation moment of $\simeq 7
\frac{\mu_{\rm{B}}}{\rm fu}$. This suggests that Europium is in the divalent
state and one would expect an insulating behavior, which could, however, not be confirmed by the corresponding 
measurements of the electrical resistivity. In a more recent work~\cite{wandner10}, the ferromagnetic order
with a somewhat smaller $T_{C} \simeq 15$K has been confirmed, but the measured electrical resistivity $\rho(T)$ 
shows an activated characteristics within the  paramagnetic phase and varies over several orders of magnitude when 
the sample is heated from $T_{C}$ to room temperature, whereas, according to Ref.~\onlinecite{sakai82}, $\rho(T)$ is 
almost temperature independent in that temperature range. Despite these very different temperature dependences of 
$\rho(T)$, both works agree at least qualitatively in the observation of a strong suppression of $\rho$ in the ferromagnetic ordered phase. The
$\rho(T)$ data of EuC$_{2}$ presented in 
Ref.~\onlinecite{wandner10} strongly resemble those of the highly investigated EuO, which may be viewed as one of the most suitable model systems 
for future spintronics devices. 

In this report, we present a detailed study of the thermodynamic properties specific heat, magnetization, and linear thermal expansion
as well as electrical resistivity measurements of various EuC$_{2}$ samples, which are prepared under slightly varying conditions. 
The analysis of the thermodynamic properties reveals that the magnetic ordering transition can be consistently described by assuming 
localized Eu$^{2+}$ moments with spin $S=7/2$ that are coupled ferromagnetically. In particular, these data do not yield any indication
that the magnetic ordering transition is related with a sizable change of the average valence state of the Eu ions. For the electrical 
transport behavior, we typically find a semiconducting behavior in the paramagnetic phase and a decrease of $\rho(T)$ in the ferromagnetically 
ordered phase. The $\rho(T)$ curves measured on different samples, however, vary with respect to the absolute value and/or the activation energy.
This sample dependence is very pronounced when the preparation conditions are changed, but -- to a less extent -- is also present for samples
which are prepared under the nominally same conditions.

\section{Experimental}

In order to synthesize EuC$_{2}$, a total amount of about 2~g of Eu and 
graphite powders were mixed in the molar ratio of $1:2.2$ in a ball mill placed
inside an Argon-filled glove box. The small surplus of graphite was used 
to inhibit the formation of EuO and to account for
graphite losses due to a reaction with the container wall. The mixed
graphite and Eu powder was transferred into a purified
Ta ampoule, which was sealed in He atmosphere (800~mbar). Then the sealed
ampoule was heated in argon atmosphere to 1400$^\circ$C with a heating rate
of 400$^\circ$C/h, kept there for 24 hours, and then cooled to room temperature with 100$^\circ$C/h. 
It is also possible to decrease the reaction temperature to 1200$^\circ$C without changing the
physical properties of the synthesized samples. In this latter case the sealed Ta ampoule was
placed inside a quartz ampoule, which was sealed in vacuum and the quartz ampoule was heated in air. The resulting EuC$_{2}$
powder is black and no impurity phases were detectable by x-ray powder
diffraction (Huber G670, Mo$K_{\alpha_{1,2}}$ radiation). 
In addition, we prepared samples with different Eu:C ratios of $1:2.2 + x$ in order to get off-stoichiometric samples.
Below we will discuss samples prepared with $x=0$, $-0.1$, and $0.3$, which are labeled as EuC$_{2}$, EuC$_{2 - \delta}$ and EuC$_{2+\delta}$.
For the measurements the powder sample obtained by the procedure described above was pressed by 2000~kg to a pellet (diameter 5 mm) 
and sintered inside a Ta ampoule at 500$^\circ$C for 96~h (heating rate: 100$^\circ$C/h; cooling rate: 20$^\circ$C/h). After sintering the quality of
the sample was again checked by x-ray powder diffraction. 
From the polycrystalline pellets we have sanded pieces to a rectangular shape of typical dimensions of about
$2\times1\times1\,$mm$^3$. As already mentioned EuC$_{2}$ very rapidly decomposes when it is in contact with air.
Thus, all sample handling was carried out in a glove box with 
inert atmosphere (Ar, 99.999\%). This box has been adapted to incorporate various sample rods with measuring cells for the specific heat, the thermal
expansion and the electrical resistivity. After the sample was mounted to the respective platform in argon atmosphere, the surrounding tube was evacuated and, for the actual measurements, was put into a $^4$He bath cryostat equipped with a 14 T magnet.
For the measurements at low temperatures (0.3--50~K) we used a $^3$He evaporator system (Heliox-VL, Oxford Instr.) with 
home-built devices for heat-capacity and thermal-expansion measurements. 
The heat-capacity device uses the adiabatic heat-pulse method. The thermal-expansion device is a dilatometer
which works with a variable plate capacitor controlled by the sample length.
The resistivity has been measured in the temperature range from 5 to 300~K  in a home-built setup that provides a temperature-variable 
sample platform in vacuum. We used a standard 4-probe technique with current and voltage
contacts made by a 2-component silver epoxy. The magnetization has been studied in a physical property
measurement system (PPMS, Quantum Design) using a vibrating sample
magnetometer in the temperature range from 2 to 300\,K in magnetic fields up to 14\,T. To keep the sample in an Ar
atmosphere during the magnetization measurements it was sealed in
a quartz capillary (Suprasil) inside the glove box.

\section{Results and discussion}

 \begin{figure}
 \includegraphics[width=8cm]{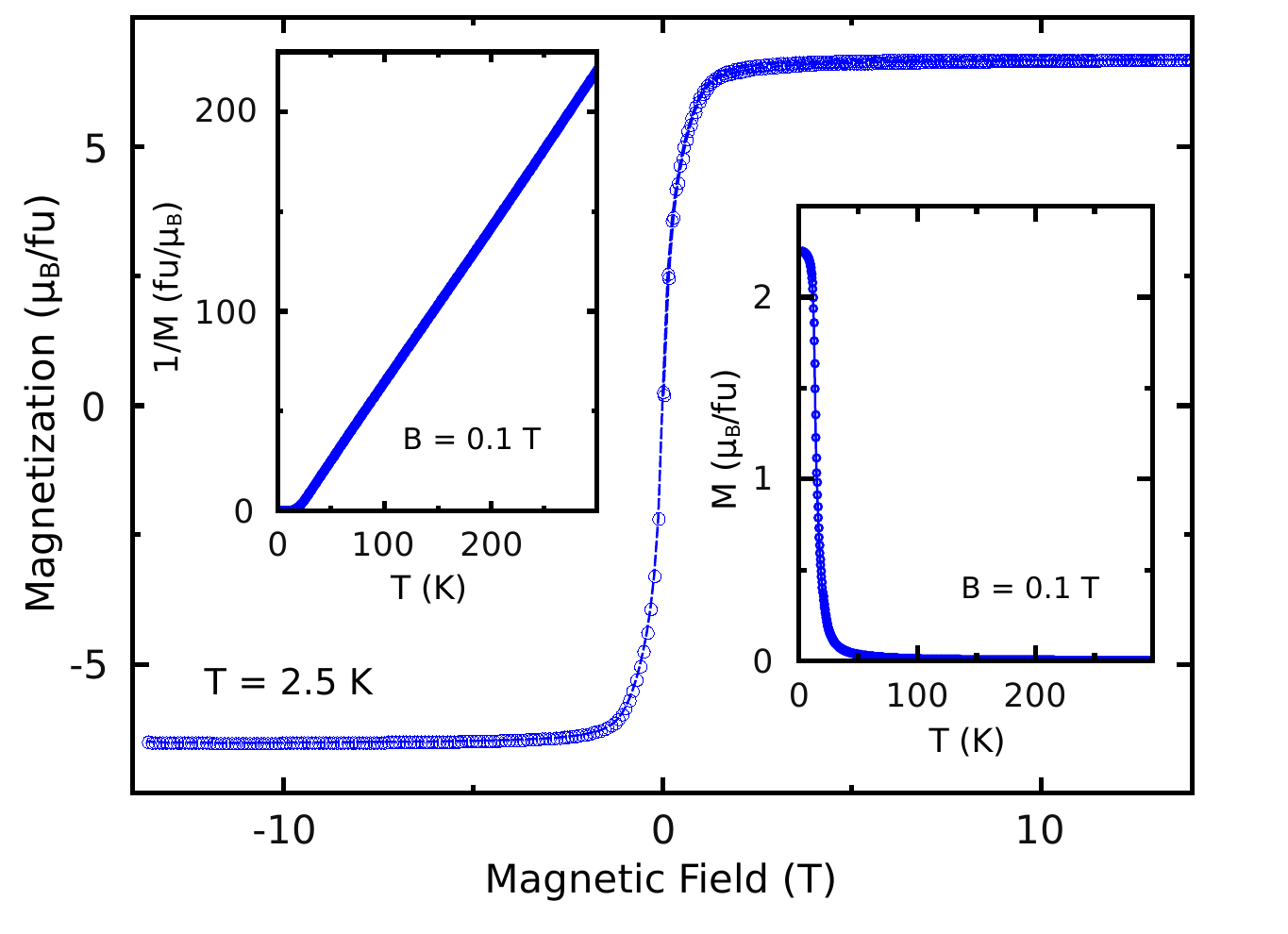}
 \caption{The field dependent magnetization of EuC$_{2}$ shows a soft ferromagnet with a negligible hysteresis and a saturation moment of
$\mu_{sat} \simeq 6.7\,\mu_{\text{B}}$. As presented in the insets, the temperature dependent $M(T)$ follows
the Curie-Weiss law in the paramagnetic phase. \label{chi}}
 \end{figure}
Fig.\,\ref{chi} displays magnetization data of EuC$_2$, which has been measured over a wide range of temperature in an external
magnetic field of $B=0.1\,$T and at 2.5\,K in the field range of $\pm 14$~T. As illustrated in the left inset of Fig.\,\ref{chi}, the $M(T)$ data are
well described by
the Curie-Weiss law $M^{-1}(T)=C^{-1}(T-\Theta)$ from 300~K down to about 20~K. Around $T_C\simeq 14$~K, $M(T)$ strongly increases and approaches an
almost constant value with further decreasing temperature. The Curie-Weiss fit yields $\Theta=17.4$\,K, i.e.\ a  ferromagnetic
exchange coupling, and an effective magnetic moment $\mu_{eff}=7.56\,\mu_{\text{B}}/$fu. The magnetization curve $M(B)$ at 2.5~K has the typical
characteristics of a soft ferromagnet with a very small hysteresis and a saturation moment of $\mu_{sat}\simeq 6.7\,\mu_{\text{B}}/$fu. Both,
$\mu_{sat}$ and $\mu_{eff}$, agree within about 4~\% to the values expected for Eu$^{2+}$-ion with $J=7/2$ ($\mu_{sat}=7$~$\mu_{\rm B}$ and
$\mu_{eff}=7.94$~$\mu_{\rm B}$). 
An important observation is that the studied sample does not show any indications of magnetic impurities. In particular, we can exclude a partial oxidation of the sample, because this would result in a
contamination with EuO, which would easily be detected in the $M(T)$ measurement because it undergoes a ferromagnetic transition at 69\,K. 

\begin{figure}
 \includegraphics[width=8cm]{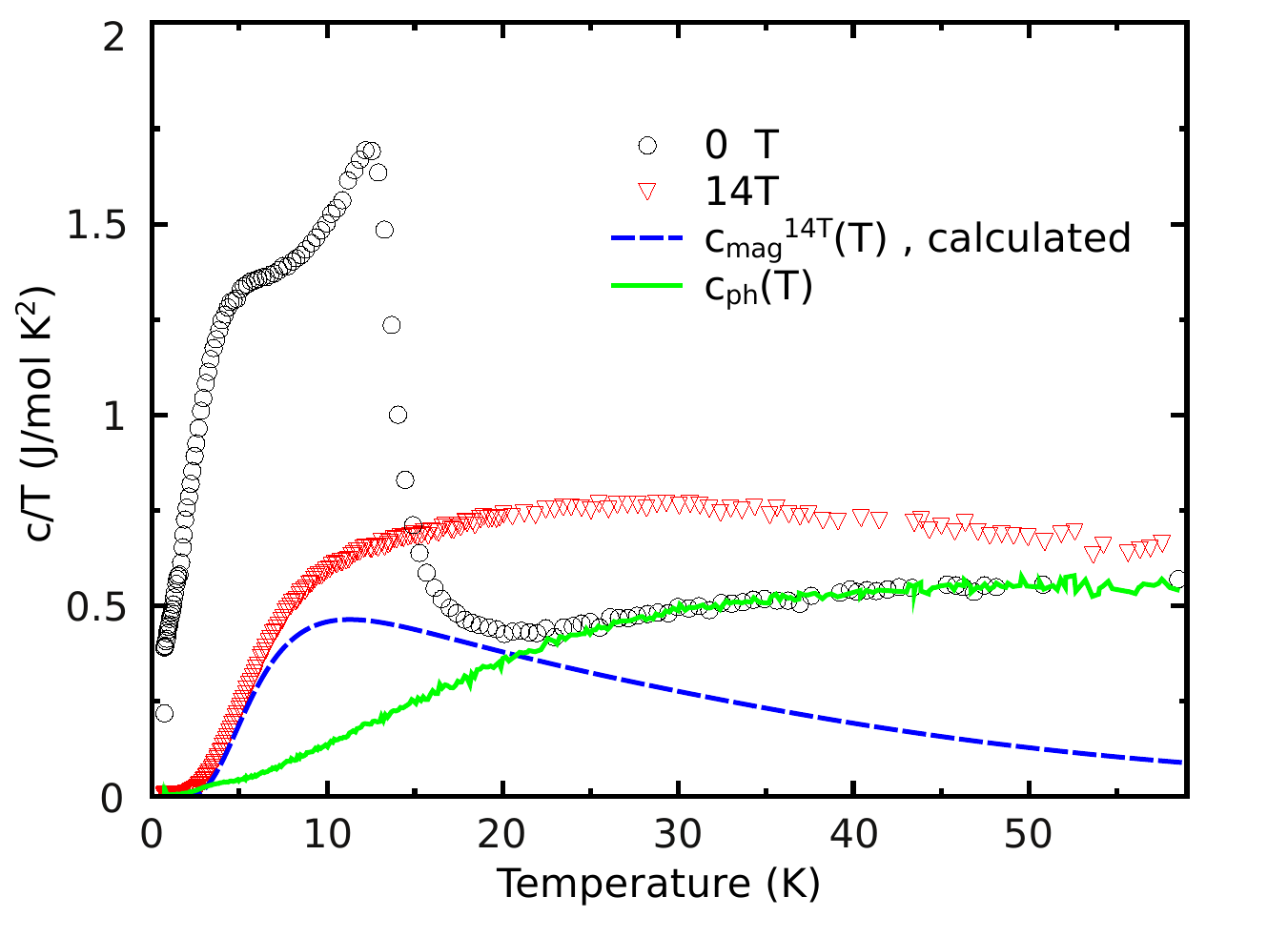}
 \caption{(Color online) Specific heat measured in zero field and in an
external magnetic field of $14$~T. The dashed line is the calculated
magnetic heat capacity of a ferromagnet in $14$~T. The
difference between the measured data and the calculated values in $14$~T shown as a solid line represents the
phonon contribution (see text).
 \label{hc}}
 \end{figure}

Fig.\,\ref{hc} displays the specific heat measurements in zero field and 14~T. The zero-field data show a pronounced anomaly at $\simeq 14$~K,
which can be attributed to the ferromagnetic ordering transition and around 4~K a broad maximum is seen in $c_p^{0T}/T$ that resembles a 
Schottky anomaly. Both features are drastically broadened in the 14~T curve as it is expected for a ferromagnet in a large magnetic field. 
For a quantitative analysis one has to separate the phononic and magnetic contributions to the total specific heat, $c_{tot}= c_{ph}+c_{mag}$. Because
we are interested in $c_{mag}$ we need an estimate of $c_{ph}$ that covers the entire temperature range up to 60~K. For this estimate we calculate
the magnon contribution  $c_{mag}^{14\,{\rm T}}(T)$ for our largest magnetic field of $14\,{\rm T}$ within a mean-field approximation and subtract it
from the experimental 
data. As the phonon contribution can be expected to be independent on the
magnetic field, this difference represents $c_{ph}$ for all applied fields. The calculated $c_{mag}^{14\,{\rm T}}/T$ is shown by the dashed line in
Fig.~\ref{hc} and the solid line displays the obtained $c_{ph}/T$, which coincides with the measured zero-field data above $\simeq 25$~K. This
appears reasonable, because the zero-field $c_{mag}^{0T}$ is expected to rapidly fall off above $T_c$. In principle, one could estimate $c_{ph}$
by subtracting a zero-field calculation of $c_{mag}^{0T}$ from the respective measured data. The mean-field approximation neglects, however, the
fluctuations around $T_c$ and the gapless spinwave excitations at low temperatures, which strongly influence the temperature dependence of $c_{mag}$ 
in zero field. Thus, we performed the mean-field calculation for a magnetic field of 14~T, which is large enough to strongly broaden the transition and to induce a sizable gap in the excitation spectrum. 

Within a mean-field approximation the internal energy per mol formula units of a Heisenberg ferromagnet 
can be expressed by \cite{rodriguez05}:
\begin{equation}
 E = - \frac{3 N_A k_B T_c S}{2 (S+1)}
\frac{M^2}{M_{sat}^2} - \mu_{\rm{B}} N_A B M
\label{emag}
\end{equation}
Here, $N_A$ and $k_B $ denote Avogadro's constant and Boltzmann's constant, respectively, $S = 1/2, 1, 3/2, ...$ is the spin number, $\mu_{\rm{B}}$
the Bohr magneton, $T_c$ the zero-field transition temperature, and $M_{sat}=2 \mu_{\rm{B}} S$ 
the saturation magnetization. Thus, the only parameter to be calculated is the 
temperature and magnetic-field dependent magnetization $M(T,B)$. As described in standard textbooks~\cite{blundell}, $M(T,B)$ follows from an implicit
equation that can be solved numerically. From equation~(\ref{emag}) the magnetic contribution to the molar heat capacity is calculated
via
\begin{equation}
 c_{mag} = \frac{dE}{dT} = - \frac{3 N_A k_B T_c SM}{(S+1) M_0^2} \frac{dM}{dT} -
\mu_{\rm{B}} N_A B \frac{dM}{dT} \, ,
\label{cmag}
\end{equation}
and the magnetic entropy is obtained via integration  $S_{mag}=\int
\frac{c_{mag}(T)}{T} dT$. The result can be used as a  
consistency check of the calculation, because $S_{mag}= N_A k_B \ln(2S+1)\simeq 17.3$~J/mol\,K has to be reached for all magnetic fields when the
integration is done to large enough temperatures. The experimental magnetic  entropy change of EuC$_2$ is obtained by the integration $\int
\frac{c_{tot}(T)-c_{ph}(T)}{T} dT$. For the zero-field data, this integration yields $S_{mag}\simeq 17.6$~J/mol\,K, which is very close to the
expected value of an $S=7/2$ system.

\begin{figure}
 \includegraphics[width=8cm]{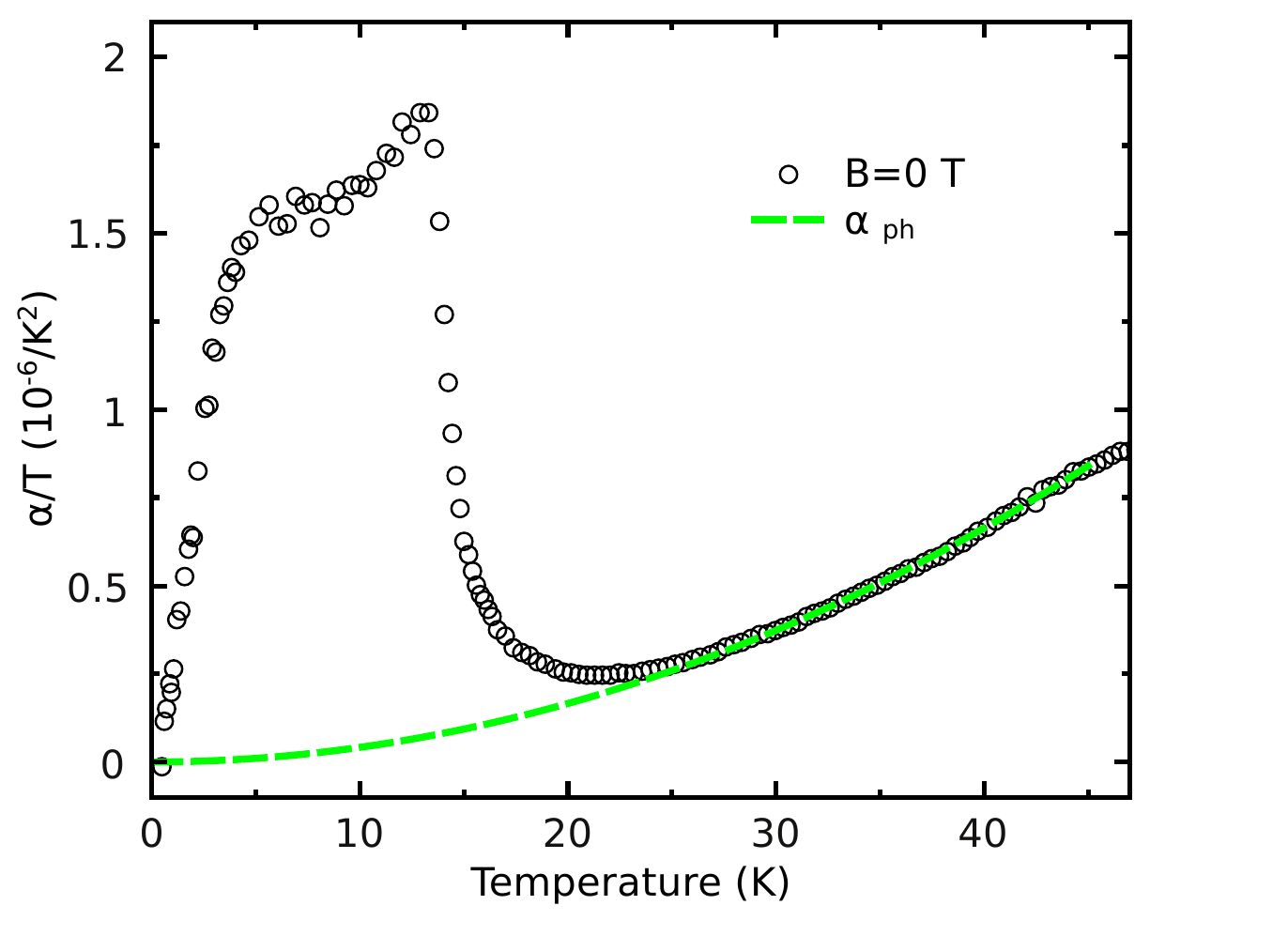} 
 \caption{The zero-field thermal expansion of EuC$_{2}$ strongly resembles the corresponding specific heat data. The phonon contribution is estimated
by a $T^3$ law $\alpha_{ph}=AT^3$ with $A=4.15 \cdot 10^{-10} \text{K}^{-4}$. 
\label{tad}}
 \end{figure}

\begin{figure}
 \includegraphics[width=8cm]{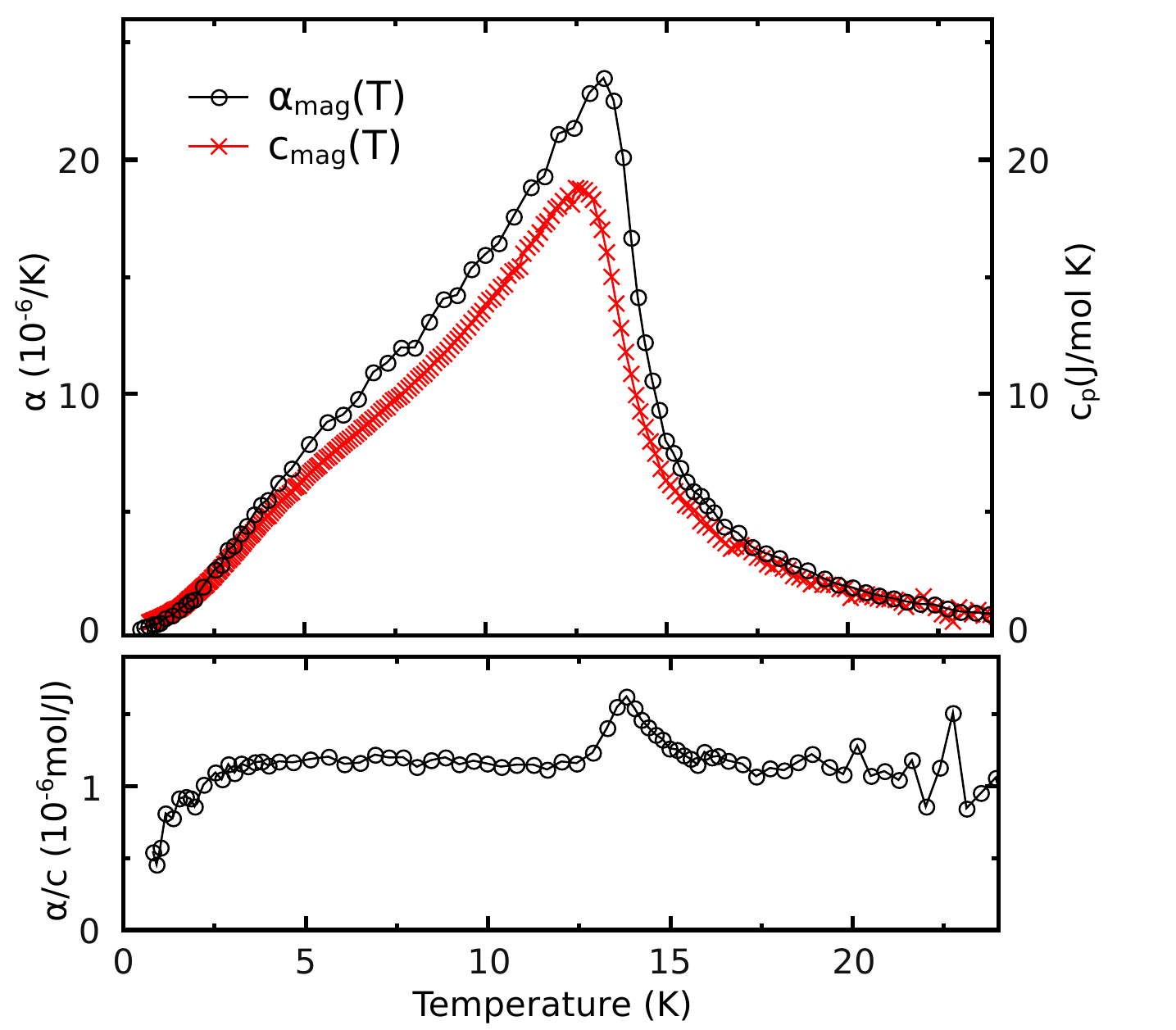}
 \caption{Magnetic contributions of the thermal expansion and specific heat (top) of EuC$_{2}$. The ratio $\alpha_{mag}/c_{mag}$ (bottom)
is essentially constant in the entire temperature as expected from a Gr\"{u}neisen scaling. \label{tadcp}}
\end{figure}

Fig.~\ref{tad} displays the zero-field linear thermal expansion of EuC$_2$ in the representation $\alpha/T$ versus $T$. Obviously, the shape of this
curve with a pronounced anomaly at $T_c\simeq 14$~K and a maximum around 4~K very strongly resembles the $c_p^{0T}/T$ data shown in Fig.~\ref{hc}. For
the further analysis we again assume the superposition of phononic and magnetic contributions $\alpha_{tot}=\alpha_{ph}+\alpha_{mag}$ and estimate
$\alpha_{ph}$ by a quadratic fit of the zero-field $\alpha_{ph}(T)/T$ data in the temperature range above 25~K (dashed line in Fig.~\ref{tad}).
An approximate proportionality between thermal expansion and specific heat is often observed in solids and is related to the Gr\"uneisen scaling,
which can be straightforwardly derived from Maxwell's relations for systems that are determined by a single energy scale~\cite{zhu03,lorenz07}. The
Heisenberg ferromagnet with nearest neighbor exchange $J$ is such a system and it has been shown explicitly~\cite{mattis63, callen65, argyle67} that
the magnetic contributions of the specific heat and thermal expansion are expected to scale with each other with a proportionality constant that is
determined by the pressure (or volume) dependence of $J$.  Thus, the magnetic contribution of the thermal expansion coefficient can be expressed as
\begin{equation}
\alpha_{mag} = \frac{\partial \ln J}{\partial p} \frac{c_{mag}}{3V_{\text{mol}}}.
\label{grueneisen}
\end{equation} 
The additional factor of 3 in the denominator takes into account that $\alpha_{mag}$ is the linear thermal expansion, which in case of a polycrystal
is 1/3 of the volume expansion. Fig.~\ref{tadcp} compares the zero-field data of $\alpha_{mag}$ and $c_{mag}$, which are obtained by subtracting
$\alpha_{ph}$ and $c_{ph}$ from the respective measurements. In the lower panel of this figure we show the ratio $\alpha_{mag}(T)/c_{mag}(T)$, which
is practically constant in the temperature range from about 2 to 22~K. The peak in the close vicinity of T$_c$  may partly arise from small
differences in
the temperature calibrations of the different experimental setups for the thermal expansion and the specific heat measurements. An additional source
for this deviation are differences in the critical behavior of $\alpha_{mag}$ and $c_{mag}$. 

The measurements of thermal expansion and specific heat also allow to derive the initial slope of the  $\partial T_c/\partial p$ at ambient pressure.
In case of a second-order 
phase transition this pressure dependence is given by the Ehrenfest relation
\begin{equation}
 \label{ehren}
 \left.\frac{\partial T_c}{\partial p}\right|_{p_0}=3 T_c V_{\text{mol}} \frac{\Delta \alpha}{\Delta c_p}\, ,
\end{equation}
where $\Delta \alpha$ and $\Delta c_p$ denote the mean-field jumps of $\alpha$ and $c_p$, respectively. 
Because there are no sharp jumps in $\alpha(T)$ and $c_p(T)$, one could approximate the experimentally 
obtained anomalies by jumps using, e.g., area-conserving constructions or it is possible to obtain a measure 
of $\partial \ln T_c /\partial p$ from a scaling of the anomalies of $\alpha(T)$ and $c_p(T)$ in the vicinity of $T_c$. This second possibility
expresses a reformulation of Eq.~(\ref{ehren}) in the form of Eq.~(\ref{grueneisen}), where $J$ is exchanged by $T_c$. Within mean-field theory $T_c$
and $J$ are proportional to each other and therefore $\partial \ln T_c /\partial p=\partial \ln J /\partial p$ meaning that Eqs.~(\ref{ehren})
and~(\ref{grueneisen}) are equivalent. Although this strict equality will not be valid in reality due to the presence of fluctuations, one may expect
an approximate equality of Eqs.~(\ref{ehren}) and~(\ref{grueneisen}). As shown in the lower panel of fig.~\ref{tadcp} the ratio
$\alpha_{mag}(T)/c_{mag}(T)\simeq 1.25\cdot10^{-6}$~mol/J over most of the temperature range and increases to $\simeq 1.5\cdot10^{-6}$~mol/J around
$T_c$. Thus, using $V_{\text{mol}}\simeq 34$~cm$^3$ we estimate the hydrostatic pressure dependences 
\begin{equation}
 \label{hydro}
 \left.\frac{\partial \ln T_c}{\partial p}\right|_{p_0}\simeq \left.\frac{\partial \ln J}{\partial p}\right|_{p_0}\simeq 0.14\,/\mbox{GPa} \, 
 \end{equation}
corresponding to an initial slope $\partial T_c / \partial p \simeq 2$~K/GPa. 
Thus, we conclude that external pressure should cause a strong increase of the ferromagnetic transition temperature of EuC$_2$ arising from a strong spin lattice coupling. 

\begin{figure}
 \includegraphics[width=8cm]{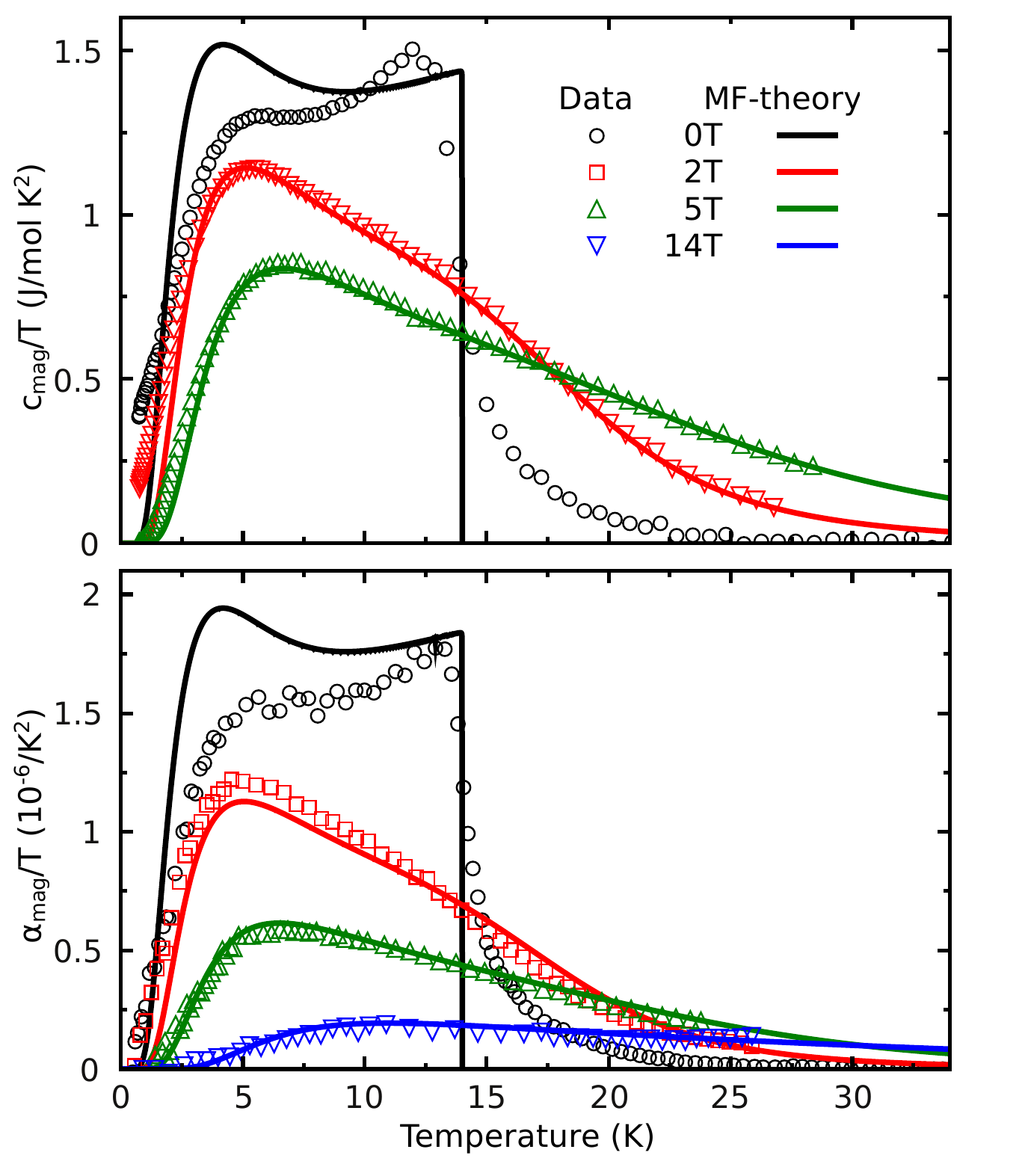}
 \caption{(Color online) Magnetic contribution of the specific heat (top) and thermal expansion (bottom) of \euc for different magnetic fields. The solid lines represent $c_{mag}$ and
$\alpha_{mag}$ calculated via Eq.~(\ref{cmag}) and Eq.~(\ref{alphamag}), respectively. 
\label{tadcpfeld}}
 \end{figure}

In the upper panel of Fig.\,\ref{tadcpfeld} the experimental magnetic contribution of the specific heat 
measured for different magnetic fields is compared to the respective $c_{mag}$ calculated via Eq.~(\ref{cmag}).
As already mentioned the mean-field calculation only yields a poor description of the zero-field data, because   it neither reproduces the spin-wave
contribution in the low-temperature region nor the behavior around $T_c$. The experimental data obtained in 2 and 5~T are, however, rather well
described, in particular in the high-temperature range. There are still deviations at low temperatures, but with increasing field these deviations
decrease. Note that the data for 14\,T is not shown here, because this measurement was used to estimate the phonon contribution $c_{mag}$. 

The lower panel of Fig.\,\ref{tadcpfeld} displays the magnetic contribution of the thermal expansion $\alpha_{mag}$ for different magnetic fields,
which behaves very similar as $c_{mag}$. For a ferromagnet with 
pressure dependent exchange interaction, mean-field theory predicts an anomalous length  
change $\Delta L/L_0$, which is proportional to the squared magnetization $M^2(T)$~\cite{callen65,chikazumi}. Thus, for the magnetic thermal
expansion $\alpha_{mag} \propto M\frac{dM}{dT}$ is expected and, in order to emphasize the close analogy between $\alpha_{mag}$ and 
$c_{mag}$, we express the magnetic thermal expansion as
\begin{equation}
\alpha_{mag}= -A \cdot \frac{3 N_Ak_B T_c SM}{(S+1)} \frac{dM}{dT} \, .
\label{alphamag}
\end{equation}
This expression is identical to the first term of Eq.~(\ref{cmag}) multiplied by a constant 
factor $A$, which is determined by the scaling behavior of the zero-field data shown in Fig.~\ref{tadcp}, i.e.~$A=\alpha_{mag}^{0T}/c_{mag}^{0T}\simeq
1.25\cdot 10^{-6}$~mol/J.  
Thus, all parameters of Eq.~(\ref{alphamag}) are fixed and $\alpha_{mag}$ calculated for different fields is shown by the solid lines in
Fig.~\ref{tadcpfeld}. Again, there are strong deviations concerning the zero-field data, whereas the agreement is satisfactory already for a field of
2~T and is getting better with further increasing field. It is interesting to note that according to the mean-field equations~(\ref{cmag})
and~(\ref{alphamag}) the scaling of the zero-field $\alpha_{mag}$ and $c_{mag}$ is no longer valid in finite magnetic fields. This mean-field result
is confirmed by the experimental data, despite the fact that the mean-field calculations do not reproduce the respective zero-field data:
Fig.~\ref{tadcp} confirms the scaling 
of the experimental zero-field data of $\alpha_{mag}^{0T}$ and $c_{mag}^{0T}$ and in Fig.~\ref{tadcpfeld} it is seen that this scaling changes in a
finite magnetic field, because the flattening of the zero-field anomaly is more pronounced for $\alpha_{mag}$ than for $c_{mag}$. 

From the thermodynamic data presented above we find that the magnetization as well as the magnetic contributions of the specific heat and the thermal expansion are satisfactorily described within a mean-field theory assuming 
Eu$^{2+}$ ions with localized $S=7/2$ moments coupled via a pressure-dependent exchange coupling $J$. In particular, these data do not show any
indications for  field- or temperature-induced changes of the Eu valence, in agreement with the results of Ref.~\onlinecite{wandner10}. M\"o{\ss}bauer
measurements done in this previous work detected a small, but temperature-independent fraction of about $4-5\,\%$ of Eu$^{3+}$. The $\simeq 4$~\%
reduction of the measured $\mu_{sat}$ and $\mu_{eff}$ compared to the expected values for an $S=7/2$ system could also be explained by such an amount
of Eu$^{3+}$, because Eu$^{3+}$ only has a weak paramagnetic van Vleck susceptibility. It is, however, unclear whether such an Eu$^{3+}$ content is an
intrinsic property of EuC$_2$, since it might also arise from a sample-dependent off stoichiometry and/or a weak oxygen contamination. 
The influence of weak off stoichiometries is also subject of intense studies in the closely related EuO, see e.g.~Ref.~\onlinecite{steeneken02}, and
it is found that the electrical resistivity can be drastically changed by weak variations of the Eu:O ratio. If \euc is considered as a purely ionic
compound composed of Eu$^{2+}$ and C$_2^{2-}$ ions, a semiconducting behavior would be expected over the entire temperature. This simple ionic
picture is, however, contradicted by the measured resistivity $\rho(T)$ showing an insulator-to-metal transition that coincides with the ferromagnetic ordering in zero-field~\cite{wandner10}. Moreover, \euc shows a giant
magneto resistance effect in finite magnetic fields with resistivity changes of several orders of magnitude around $T_c$. Thus, \euc behaves very similar to EuO. 

\begin{figure}
 \includegraphics[width=8cm]{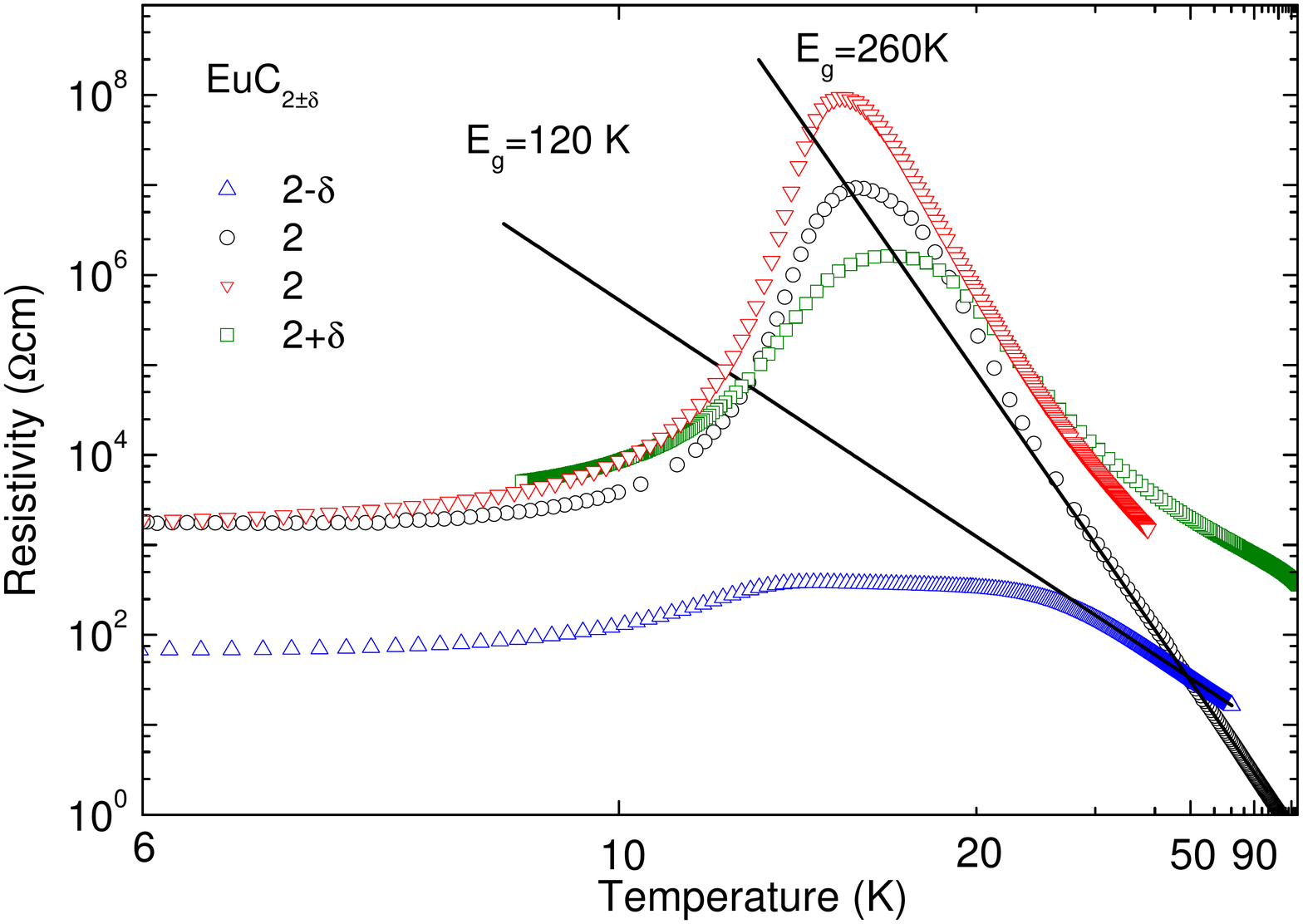}
 \caption{(Color online) Arrhenius plot of the resistivity of various EuC$_{2\pm \delta}$ samples prepared for different E:C ratios  (see experimental section). The black lines represent the curves for constant activation
energies $E_g$. It is noticeable that the off-stoichiometric samples have a considerably smaller $E_g$.  \label{resistivityplot}}
 \end{figure}

In order to get more insight about the influence of stoichiometry, the synthesis of EuC$_{2 \pm \delta}$ samples has been performed.
Fig.~\ref{resistivityplot} compares the resistivity of two nominally stoichiometric EuC$_{2}$ samples and two off-stoichiometric samples that are expected to be either Eu- or C$_{2}$-deficient, i.e.~nominally electron- or hole-doped. The qualitative behavior for all samples is rather similar. 
Upon decreasing the temperature down to $T_c$, the resistivity continuously increases, then $\rho(T)$ strongly
decreases below $T_c$ and finally becomes practically temperature-independent below about 7~K. Based on the opposite temperature dependences above and below $T_c$, one may discuss this as an insulator-to-metal transition, but in view of the rather large absolute values of $\rho$ below $T_c$, it is also appropriate to consider an insulator-to-insulator transition. Concerning the maxima, the nominally 
stoichiometric samples reach $\rho \gtrsim 10^7\,\Omega$cm. The maximum for the nominally Eu-deficient sample 
is still larger than $10^6\,\Omega$cm, while in the C$_2$-deficient sample a broadened plateau is observed with $\rho(T)\simeq 10^2\,\Omega$cm in the
temperature range from about 12 to 20~K. Moreover, the low-temperature resistivity of EuC$_{2-\delta}$ is about one order of magnitude lower than
$\rho$ of the other samples.  In the temperature range above $30$~K, the resistivity of all samples decreases more or less linearly in the 
Arrhenius plot, i.e., there is an activated behavior $\rho(T)\propto \exp(-E_g/k_BT)$ with a temperature-independent activation energy
$E_g$. It is remarkable that both off-stoichiometric samples have a similar  $E_g \approx 120$\,K in the paramagnetic phase, whereas the
stoichiometric samples have a significantly larger activation energy $E_g \approx 260\,$K. This suggests that variations in the stoichiometry 
induce donator or acceptor levels in the bandgap, which may strongly influence the resistivity behavior, but the influence on the ferromagnetic 
ordering temperature is comparatively weak. At the present stage of sample preparation and characterization it is not possible to give a precise
determination of the real composition of the studied sample, which would be necessary to estimate the charge-carrier content. This together with the
fact that no single-crystalline samples are available, prevents a deeper analysis of the complex resistivity behavior of the different samples shown
in 
Fig.~\ref{resistivityplot}. In many aspects the data of EuC$_{2\pm \delta}$ resemble the early investigations of the Eu chalcogenides 
EuS and EuO~\cite{shapira72,shapira73a,oliver72,mauger80}. There, it was also observed that, depending on the exact preparation
process and/or subsequent tempering procedures, the resistivity may vary over several orders of magnitude, in particular in the temperature range of
the ferromagnetic ordering temperature. This has been attributed to the influence of donator and/or acceptor levels, that are controlled by the exact
stoichiometry. The occurrence of an insulator-to-metal transition at the ferromagnetic ordering of EuO is explained by an exchange splitting of 
the conduction band into spin-up and spin-down bands, which is large enough to drop one of these bands below the donator levels~\cite{steeneken02}. 
This means, however, that a finite amount of donator levels is required in order to make the ferromagnetic state conducting. Because our data of \euc 
in most aspects strongly resemble the corresponding data of EuO, we conclude that \euc represents a low-temperature analogue of EuO.

\section{Summary}
We have investigated the magnetization, specific heat, thermal expansion, and resistivity of high-quality
EuC$_2$ samples. 
The temperature and field dependence of the magnetization fit the magnetic moment expected for Eu$^{2+}$ ions with $S=7/2$. 
The specific heat $c_p(T)$ and the thermal expansion $\alpha(T)$ show well-defined anomalies at the magnetic phase transition and
the magnetic contributions of $c_{mag}$ and $\alpha_{mag}$ could be extracted for further analyses. Using  Ehrenfest's relation
to determine the pressure dependence of the Curie-temperature, an initial slope of $\partial T_c/\partial p \simeq 2\,$K/GPa is found. 
The zero-field data of $c_{mag}$ yield a magnetic entropy $S_{mag} \simeq 17.6$~J/mol\,K, which is close to the expected
value of an $S=7/2$ system. The ratio $\alpha_{mag}/c_{mag}$ is practically constant in the temperature from about 2 to 22~K meaning that the respective zero-field data obey a magnetic Gr\"uneisen scaling. This Gr\"{u}neisen scaling is, however, not preserved in finite magnetic fields.
Comparing the experimentally obtained magnetic contributions $c_{mag}$ and $\alpha_{mag}$ to the respective
results calculated within a mean-field model for an $S=7/2$ system, we find that the experimental data for larger magnetic fields are well reproduced 
by the mean-field model, whereas the calculations for zero-field only yield a poor description of the experimental data. Nevertheless, the 
mean-field model correctly predicts that the zero-field Gr\"{u}neisen scaling is no longer valid in finite fields. The influence of weak off stoichiometries was studied by comparing the resistivity measurements of nominally
stoichiometric EuC$_2$ and off-stoichiometric samples, which are expected to be either Eu or C$_2$ deficient. The variations in the stoichiometry most probably induce donator or acceptor levels in the bandgap, which strongly affects the resistivity behavior, whereas the influence on
$T_C$ is comparatively weak. In all these aspects the resistivity data strongly resemble the early investigations of the Eu
chalcogenide EuO. 

\begin{acknowledgments}
This work was supported by the Deutsche Forschungsgemeinschaft through
Schwerpunktprogramm~1166 and Sonderforschungsbereich~608.
\end{acknowledgments}


\end{document}